\documentstyle[12pt]{article}

\newcommand{\eq}{\begin{equation}}
\newcommand{\en}{\end{equation}}
\newcommand{\bea}{\begin{eqnarray}}
\newcommand{\eea}{\end{eqnarray}}
\newcommand{\spz}{\hspace{0.7cm}}
\newcommand{\virg}{\spz,\spz}

\newcommand{\half}{{\textstyle\frac{1}{2}}}
\newcommand{\sfrac}[2]{{\textstyle\frac{#1}{#2}}}
\newcommand{\im}{\Im m}

\newcommand{\intr}{\int_{-\infty}^{+\infty}}
\newcommand{\ct}{\tilde{c}}

\newcommand{\bZ}{\mbox{\bf Z}}

\newcommand{\cH}{{\cal H}}
\newcommand{\cI}{{\cal I}}

\newcommand{\NP}[1]{Nucl.\ Phys.\ {\bf #1}}
\newcommand{\PL}[1]{Phys.\ Lett.\ {\bf #1}}

\newcommand{\PRL}[1]{Phys.\ Rev.\ Lett.\ {\bf #1}}

\newcommand{\IJMP}[1]{Int.\ J.\ Mod.\ Phys.\ {\bf #1}}

\begin{document}
\sloppy
\renewcommand{\thefootnote}{\fnsymbol{footnote}}

\newpage
\setcounter{page}{1}

\vspace{0.7cm}
\begin{flushright}
DFUB 96-24\\
August 1996\\
hep-th/9608091
\end{flushright}
\vspace*{1cm}
\begin{center}
{\bf EXCITED STATE DESTRI -- DE VEGA EQUATION FOR SINE--GORDON AND
RESTRICTED SINE--GORDON MODELS}\\
\vspace{1.8cm}
{\large D.\ Fioravanti, A.\ Mariottini, E.\ Quattrini and F.\ Ravanini}\\
\vspace{.5cm}
{\em Sez. I.N.F.N. and Dip. di Fisica - Univ. di Bologna\\
     Via Irnerio 46, I-40126 BOLOGNA, Italy}
\end{center}
\vspace{1cm}

\renewcommand{\thefootnote}{\arabic{footnote}}
\setcounter{footnote}{0}
 
\begin{abstract}
We derive a generalization of the Destri - De Vega equation governing 
the scaling functions of some excited states in the Sine-Gordon
theory. In particular configurations with an even number of holes and
no strings are analyzed and their UV limits found to match some of the
conformal dimensions of the corresponding compactified massless free boson.
Quantum group reduction allows to interpret some of our results as 
scaling functions of
excited states of Restricted Sine-Gordon theory, i.e. minimal models
perturbed by $\phi_{13}$ in their massive regime. In particular we are
able to reconstruct the scaling functions of the off-critical
deformations of all the scalar primary states on the diagonal of the Kac-table.
\end{abstract}
\newpage

\section{Introduction}
The exact computation of scaling functions in two-dimensional integrable
quantum field theories has been paid a lot of attention in the last
years. The Thermodynamic Bethe Ansatz (TBA) methods~\cite{Al1,YY}, 
for example, are able
to provide, for a theory whose factorizable S-matrix is known, a very
good framework to compute a set of coupled non-linear integral
equations driving the evolution along the Renormalization
Group flow of the Casimir energy of the vacuum on a cylinder. Towards
the UV limit this quantity is directly related to the central charge
of the corresponding Conformal Field Theory.

However, the TBA method suffers of some disadvantages. First of all the
deduction of the integral equations is done starting from the
Factorized Scattering Theory (FST) and not from the Quantum Field Theory
(QFT) itself. Now, the FST is often only {\em conjectured} and the TBA
itself is therefore not fully {\em deduced} from the QFT. Another
point is that there are models where the number of equations for the
TBA becomes infinite or increases to a point that numerical integration 
becomes untractable.

An undoubtable progress of these exact methods came with the introduction, at
least for the Sine-Gordon model, of
the so called Destri -- De Vega (DdV) equation~\cite{DdV1,DdV2}, 
whose major advantages are:
\begin{itemize}
\item to summarize in one complex integral equation a large number, sometimes
an infinity, of real coupled integral equations of TBA type;
\item to be deduced from a lattice regularization of the model itself
and not from the scattering theory corresponding to it.
\end{itemize}
Unfortunately the DdV equation is developed and well explored only for
the vacuum of the Sine-Gordon model and it would be very important 
to write down
analogous equations for other integrable field theories too, as well
as for excited states.
In this letter we make progress towards the study of DdV equation for
excited states in the Sine-Gordon model. The quantum group reduction
allows then to extend these results to the Restricted Sine-Gordon
models, i.e. the integrable theories obtained by perturbing the conformal
minimal models by their relevant operator $\phi_{13}$. This is the
first case of extension of the DdV equation out of the pure
Sine-Gordon realm.

In sect.2, after recalling some general properties of the SG model on
a cylinder, we write down, following~\cite{DdV2,Zam-polymers}, the DdV
equation for vacua with twisted boundary conditions. These vacua are
states that can be obtained from the true vacuum by application of {\em
spin fields}. In sect.3 we observe that after quantum group reduction,
some of these states can be reinterpreted in the framework of RSG
theory as the off-critical deformations of scalar primary fields along
the main diagonal of the Kac-table. A very precise numerical check
against known cases confirms this procedure. In sect.4 we afford the
problem of writing the DdV equation for excited states over the
twisted vacua, by exciting Bethe strings and
creating holes over the Dirac sea of real solutions.
This corresponds to approaching the problem in a true Bethe
ansatz spirit. In this short letter we shall make the choice to
illustrate for technical simplicity only the case of pure {\em hole}
excitations, leaving the more general problem of string excitations to
a future, more extensive publication~\cite{preparation}. In sect.5 we
compute the energy as a functional of the solution of the DdV
equation. In sect.6 we investigate the ultraviolet (UV) limit of the
energy and compare with the operator content of the compactified free
boson CFT that represents the UV limit of SG theory. Some preliminary
numerical investigations are reported in sect.7 and in sect.8 we draw
our conclusions and perspectives for future work. 

\section{The Sine-Gordon Hilbert Space on a cylinder}

The Sine-Gordon (SG) model, with action
\eq
S = \int d^2x \left\{ \half :(\partial \phi)^2: + \lambda :\cos 
\beta\phi:\right\}
\en
is invariant under the symmetry $\phi\to\phi+2\pi n/\beta$. 
States transforming as $|\Psi\rangle\to e^{i\alpha}|\Psi\rangle$ under
this symmetry constitute a so-called $\alpha$-sector $\cH_{\alpha}$ of
the Hilbert Space of states $\cH = \oplus_{-\pi<\alpha\leq\pi}
\cH_{\alpha}$. On a cylinder of radius $R$ the degeneracy among the 
$\alpha$-vacua is removed and the various $\alpha$-sectors are selected
by imposing suitable twisted boundary conditions on the space direction
of the cylinder.
For $p=\frac{\beta^2}{8\pi-\beta^2}$ rational, the
values of $\alpha$ are quantized; if $p$ is integer, i.e. the case
that we shall mainly consider in the following, then
$\alpha=\frac{\pi \kappa}{p}$, $\kappa=0,1,...,p$. The $\alpha$-sectors have of
course a state of lowest energy, the so called $\alpha$-vacuum. For
$\alpha=0$ this is the true vacuum of the theory, the other
$\alpha$-vacua can be excited from the true vacuum by particular
operators which are called {\em spin operators}. In the CFT
representing the ultraviolet (UV) limit of SG at a fixed $p$ (SG$_p$) 
(a $c=1$ massless free
boson compactified on a circle of radius $\frac{\sqrt{4\pi}}{\beta}$) these
operators are represented by vertex operators of conformal dimensions
$\Delta_{\kappa} = \frac{\kappa^2}{4p(p+1)}$. For more details see, e.g., 
ref.~\cite{Zam-painleve}.

A DdV equation can easily be written for all the $\alpha$-vacua along
the lines of the standard deduction of the original work of Destri and
De Vega~\cite{DdV1,DdV2}. Indeed, one simply has to consider twisted
boundary conditions, for which a DdV equation is written
in~\cite{DdV2} for the XXZ chain. It is a straightforward exercise to
rewrite it in the form more appropriate for the SG$_p$ model. The
resulting equation is~\cite{Zam-polymers}
\eq
f_{\alpha}(\theta) = i r \sinh \theta + i \alpha + 2i
\int_{-\infty}^{+\infty} d\theta'
G(\theta-\theta') \im \log (1+e^{f_{\alpha}(\theta+i0)})
\en
where $r=mR$, $m$ being the soliton mass. The kernel 
$G(\theta)=-\frac{i}{2\pi}\frac{d}{d\theta}\log S(\theta)$ is
determined by the soliton-soliton scattering amplitude
\eq
S(\theta)=\exp\frac{i}{2} \intr dk \frac{\sin(\theta k)}{k} 
\frac{\sinh\frac{k\pi}{2}(p-1)}{\sinh\frac{kp\pi}{2}\cosh\frac{k\pi}{2}}
\en
Once the pseudoenergy $f_{\alpha}(\theta)$ is determined by solving the DdV
equation, it can be used to compute the ground state Casimir energy
$E_{\alpha}(R)=-\frac{\pi c_{\alpha}(r)}{6R}$, where
\eq
c_{\alpha}(r)=\frac{3r}{\pi^2}\intr d\theta \sinh\theta
\im\log(1+e^{f_{\alpha}(\theta+i0)})
\en
The adimensional function $c_{\alpha}(r)$
is known as scaling function for the $\alpha$-sector ground state.
A standard dilogarithm calculation shows that the UV limit $r\to 0$ of
this function is given by
\eq
c_{\alpha}(0)=c_{UV}-12(\Delta_+-\Delta_-)=
1-\frac{6p}{p+1}\left(\frac{\alpha}{\pi}\right)^2
\en
thus reproducing the correct central charge $c=1$ of the free boson and
the correct conformal dimensions $\Delta_+$ and $\Delta_-$ 
of the spin operators mentioned before.

One can thus follow the renormalization group destiny
of the $\alpha$-vacua in the different sectors on the cylinder in an
exact way, overcoming the need of approximante methods as the
Truncated Conformal Space Approach (TCSA)~\cite{YZ}. The obvious
result that one can see from a plot of the functions $c_{\alpha}(r)$
for all $\alpha$'s is that they all accumulate towards infrared to the
vacuum of the theory.

\section{RSG$_p$ spin states}

An interesting application of this result is the possibility to
reinterpret it in the light of the quantum group truncation
occourring at rational $p$, which,  as shown by~\cite{Smirnov,Bernard-LeClair},
selects inside the Hilbert space of the full SG$_p$ model,
a certain subspace with a consistent subalgebra of operators
acting on it in such a way that it can be thought as a sort of smaller
theory embedded in the larger SG$_p$ one. The SG vacuum
never appears as a state in this {\em restricted} Sine-Gordon (RSG$_p$)
model. It
is known that the role of the vacuum is instead played in this context
by the $\alpha$-vacuum with $\kappa=1$ (we restrict here for simplicity
to the integer $p$ case, it is not difficult to generalize to any
rational $p$). Indeed, $\lim_{r\to
0} c_{\pi/p}(r) = 1-\frac{6}{p(p+1)}$, as it must be for 
the correct scaling function of the vacuum of the corresponding
minimal model perturbed by $\phi_{1,3}$. To support this result we made
numerical comparison of the scaling function obtained by this DdV
equation with that obtained from the traditional TBA of ref.~\cite{Al3}.
The two functions agree with 14 significant digits, for integer values
of $p$ ranging from 3 to 12.

The quantum group truncation eliminates only two $\alpha$-vacua: the
$\kappa=0,\pi$ ones. All the other $\alpha$-vacua have an interpretation
as RSG$_p$ states, namely they correspond to the
(off-critical evolution of) the primary states on the diagonal of the
Kac-table of the $p$-th minimal model, the so called {\em spin states}. 
Indeed the computation of the UV
limit of the scaling function gives in general
$c_{\kappa\pi/p}(0)=1-\frac{6}{p(p+1)}-24\frac{\kappa^2-1}{4p(p+1)}$,
thus providing the conformal dimensions of the states in the diagonal of
the Kac-table. The scaling functions of a few of these states were
already explored by other
authors with other 
means~\cite{Fendley,KM-excited,Martins-excited}. We succeded in various
cases to compare our numerical results with theirs, getting the same
kind of good agreement up to 14 significant digits as for the vacuum.

The quantum group reduction for other states, especially for
secondaries, is more involved and a full treatment goes out of the
scope of the present letter. Let us only comment here that
unfortunately none of the hole excitations we are going to describe in
the following is in the class (type II representations) that survives
quantum group reduction. 

\section{DdV for pure hole excited states}

In this section we present our derivation of the DdV equation
describing {\em pure hole} excitations obtained from the vacuum 
by creating some holes in the distribution of real roots.

Let us consider the Bethe equations for the SG$_p$ model discretized 
as a 6-vertex model on a light-cone lattice with $2N$ sites. Also, take
the corresponding inhomogeneous spin chain have $\omega$-twisted
boundary conditions, related to the previous $\alpha$ by
$\alpha=-\omega\frac{p+1}{2p}$. These Bethe equations have the form
\eq
\prod_{m=1}^M \frac{\sinh (\lambda_j - \lambda_m + i\gamma)}
{\sinh (\lambda_j - \lambda_m - i\gamma)} =
- \prod_{n=1}^{2N}
\frac{\sinh (\lambda_j + i\Theta_n + \sfrac{i\gamma}{2})}
{\sinh (\lambda_j + i\Theta_n - \sfrac{i\gamma}{2})}
e^{i\omega}
\en
Here $\gamma=\frac{\pi}{p+1}$. 
The fundamental state ($\alpha$-vacuum) 
is the one with $M=N$. The light-cone approach
corresponds to the choice of inhomogenuities 
$i\Theta_n=(-1)^{n+1}\Theta$. Define the
function
\eq
\phi(\lambda,x)=i\log\frac{\sinh(ix+\lambda)}{\sinh(ix-\lambda)}
\en
and the $M$ real solutions counting function
\eq
Z_M(\lambda) = N [\phi(\lambda+\Theta,\sfrac{\gamma}{2})+
\phi(\lambda-\Theta,\sfrac{\gamma}{2})] - \sum_{k=1}^M
\phi(\lambda-\lambda_k,\gamma) - \omega
\en
Then, the logarithm of the Bethe equations can be written 
as\footnote{In this paper we are interested only in states with $M$
even}
\eq
Z_M(\lambda_j) = 2\pi I_j \virg I_j \in \bZ + \half \virg j=1,...,M
\en
We consider here a configuration with $M<N$ real solutions and $H$ holes 
(i.e. $H$ values of $I_j$ are missing).
Consistency of the Bethe equations requires that the number $H$ must be 
even~\cite{DeVega,Kar}. 
The range of possible values of $I_j$ actually increases due to this 
``perturbation'' of the system. Now there are in total $T=M+H=N+\frac{H}{2}$
possibilities, while the roots are $M=N-\frac{H}{2}$. In order to implement
the usual DdV trick, we have to rewrite the counting function $Z_M(\lambda)$
by adding and subtracting the contribution of holes
\eq
Z_M(\lambda)=NQ(\lambda) -\omega - \sum_{k=1}^T
\phi(\lambda-\lambda_k,\gamma) +
\sum_{h=1}^H \phi(\lambda-\lambda_h,\gamma) 
\en
where we have indicated for short 
$Q(\lambda)=\phi(\lambda+\Theta,\frac{\gamma}{2}) 
+\phi(\lambda-\Theta,\frac{\gamma}{2})$.
The standard derivation of DdV eqautions makes use of the trick
\eq
\phi(\lambda-\lambda_j,\gamma) = \oint_{\lambda_j} \frac{d\mu}{2\pi i}
\phi(\lambda-\mu)\frac{d}{d\mu}\log (1+ e^{i Z_M(\mu)})
\en
to write (the integration path $\Gamma$ contours the whole real axis 
counterclockwise)
\bea
\sum_{k=1}^T \phi(\lambda-\lambda_k,\gamma) &=&
\oint_{\Gamma}\frac{d\mu}{2\pi i} \phi(\lambda-\mu)\frac{d}{d\mu}\log(1+
e^{i Z_M(\mu)})\\
&=& i(X*L)(\lambda) + i(X*Z_M)(\lambda)
\eea
where $L(\lambda)=\log(1+e^{i Z_M(\lambda+i\eta)}) - 
\log(1+e^{-i Z_M(\lambda-i\eta)})$ for $\eta\to 0$, * denotes standard
convolution $(f*g)(x)=\intr 
f(x-y)g(y)dy$ and we have introduced the notation
$X(\lambda)=\frac{1}{2\pi}\phi'(\lambda,\gamma)$.

The rest of the derivation repeats the same steps as the standard
one for the vacuum DdV~\cite{DdV1,DdV2},
with the only modification that $NQ(\lambda)$ must be now substituted with
$A \equiv NQ - \omega + \sum_{h=1}^H \phi(\lambda-\lambda_h,\gamma)$.
This goes to modify the
$imR\sinh \theta$ term because instead of computing $N(K*Q)(\lambda)$ we have
to compute now $(K*A)(\lambda)$, where $K(\lambda)$ 
is the inverse of $(\delta + X)(\lambda)$, that is
\eq
[K*(\delta + X)](\lambda) = \delta(\lambda)
\en
We have to compute
terms like $K*\phi(\lambda-\mu,\gamma)$. 
Just doing a Fourier transform we get
in the case of $\mu$ real:
\eq
K\ast \phi(\lambda-\mu,\gamma) = -i\log S(\sfrac{\pi}{\gamma}(\lambda-\mu))
\en
All these calcualtons are easily performed in Fourier transformed space,
by remembering that the Fourier Transform of
$\phi(\lambda,\gamma)$ is given by $\tilde{\phi}(k,\gamma)= 
\mbox{\rm sign}(\gamma) 
\frac{2\pi}{ik}\frac{\sinh(k(\frac{\pi}{2}-|\gamma|))}{\sinh
\frac{k\pi}{2}}$.

The continuum form of DdV equation is recovered if we let
$\theta=\lambda\pi/\gamma$, $\alpha=-\omega\frac{p+1}{2p}$ 
and $f(\theta)=\lim_{N\to\infty} i Z_M(\lambda)$
\bea
f(\theta) &=& imR\sinh\theta+i\alpha
+\sum_{h=1}^H \log S(\theta-\theta_h) \nonumber \\
&+& 2i \int_{-\infty}^{+\infty} d\theta' G(\theta-\theta') \Im m
\log(1+e^{f(\theta'+i\eta)})
\label{ddv}
\eea
where $\theta_h=\frac{\pi}{\gamma}\lambda_h$

The condition we use to determine the parameters $\theta_h$ is the fact that
they are {\em missing roots}, i.e. although they are not taken as roots of
the Bethe equations, nevertheless they satisfy $Z_M(\lambda_h)=2\pi I_h$ for
some chosen $I_h$ (the set of $I_h$ for $h=1,...,H$ then labels the excited
state we are examining).
In the continuum limit this condition reads as
$f(\theta_h)=2i\pi I_h$. To numerically integrate the DdV
equation we start from the vacuum evaluation of $f(\theta)$ and search
for values in the $\theta$ complex plane where $f(\theta)=2 \pi I_h$.
We then insert 
this rough determination of $\theta_h$'s
in the DdV equation, and iterate to get a better determination of $f$.
With this better determination we recompute the values of $\theta_h$
and then back to the DdV for a newer determination of $f$ and so on
iteratively. Although the convergence of this procedure slows down if
compared to that of the vacuum case, it is still possible to
determine the functions without much effort to some 7 or 8 significant digits. 

\section{The energy}

The energy of the state
as a functional of $f$ can be obtained as the logarithm of the 
eigenvalue of the diagonal to diagonal 
transfer matrix corresponding to the state at exam
\eq
E_M = \frac{N}{R}\sum_{j=1}^M [\Phi(\lambda_j) - 2\pi]
\en
where $\Phi(\lambda)=\phi(\lambda+\Theta,\frac{\gamma}{2})-
\phi(\lambda-\Theta,\frac{\gamma}{2})$. As before, we add and subtract a 
contribution from the holes
\eq
E_M = \frac{N}{R}[\sum_{j=1}^T \Phi(\lambda_j) - \sum_{h=1}^H \Phi(\lambda_h)
-2\pi M]
\en
The first sum in this expression can be manpulated in a manner similar to 
what done in the deduction of DdV equation
\bea
\sum_{j=1}^T \Phi(\lambda_j) &=& \oint_{\Gamma}\frac{d\mu}{2\pi i} \Phi(\mu)
\frac{d}{d\mu}\log(1+e^{iZ_M(\mu)}) \nonumber \\
&=& -\int_{-\infty}^{+\infty}\frac{dx}{2\pi} \Phi'(x)Z_M(x) -
i\int_{-\infty}^{+\infty}\frac{dx}{2\pi}\Phi'(x+i\eta)
\log(1+e^{iZ_M(x+i\eta)}) \nonumber \\
&+&i\int_{-\infty}^{+\infty}\frac{dx}{2\pi}\Phi'(x-i\eta)
\log(1+e^{iZ_M(x-i\eta)})
\eea
Now we substitute the $Z_M(\mu)$ appearing in the first integral of 
this expression using the DdV equation (\ref{ddv}). Also, rewrite the
last term $2\pi M$ as $2\pi (N-\frac{H}{2})$. Introducing the
notations
\eq
E_c = \frac{N^2}{R}\left( -2\pi + \intr d\lambda
\frac{\phi(\lambda+2\Theta,\sfrac{\gamma}{2})}
{\gamma\cosh\sfrac{\pi}{\gamma}\lambda} \right)
\en
and
\eq
\psi(\lambda) = \frac{N}{\gamma R}[\mbox{\rm sech}
(\sfrac{\pi}{\gamma}(\lambda-\Theta))
- \mbox{\rm sech}(\sfrac{\pi}{\gamma}(\lambda+\Theta))]
\en
we can now write the energy as follows
\bea
E_M &=& E_c -i \intr
d\lambda\psi(\lambda+i\eta)\log(1+e^{iZ_M(\lambda+\i\eta)})
\nonumber \\
&+& i \intr d\lambda\psi(\lambda-i\eta)\log(1+e^{-iZ_M(\lambda-\i\eta)})
\nonumber \\
&+& i \frac{N}{R}\sum_{h=1}^H \intr \frac{dx}{2\pi} \Phi'(x)\log
S(\sfrac{\pi}{\gamma} (x-\lambda_h)) + \frac{N}{R}H\pi -
\frac{N}{R}\sum_{h=1}^H \Phi(\lambda_h)
\eea
The first term $E_c$ reproduces the standard bulk
energy~\cite{DdV2}. The second term, in the $N\to\infty$ limit
acquires the same form as the vacuum energy in
ref.~\cite{DdV2}. Notice, however, that this similarity is only
formal, because now this term is evaluated with the ``excited''
$f(\theta)$. The physical interpretation is that of an energy of the
vacuum distorted by the presence of particle excitations. The last
three terms are new and, when evaluated in the $N\to\infty$,
$\Theta=\frac{\gamma}{\pi} \log\frac{4N}{r} \to\infty$ limit, concur
to give, after some algebra, a term of the kind $m\sum_{h=1}^H
\cosh\theta_h$, with the clear interpretation as a particle
energy. Concluding, the continuum limit energy $\tilde{E}(R)$ 
of the excited state, epurated from the bulk
energy and the infinities which are artifacts of the lattice, is
\eq
\tilde{E}(R) = \lim_{N\to\infty}(E_M - E_c) = m \sum_{h=1}^H \cosh \theta_h -
\frac{m}{\pi} \intr d\theta \sinh\theta \im \log (1+e^{f(\theta+i0)})
\en
Note that in the infrared limit
$r\to\infty$, where the convolution integral can be dropped, an
$H$-hole state can be interpreted as an $H$-particle state. Having in
our approach $H$ even only, we see states with an even number of
particles only. To deal with the more general problem of any number of
particles, one should also consider DdV equations which
are continuum limits over lattices with odd number of sites, or
equivalently choose twisted boundary conditions in an appropriate
way. This is, however, out of the scope of the present letter.

\section{Kink DdV equations and analytic ultraviolet limit}

The scaling function $\ct(r)=-6R\tilde{E}(R)/\pi$ is particularly
interesting because, in the ultraviolet (UV) 
limit $r\to 0$, it provides the value of the
conformal dimensions $\Delta_++\Delta_-$ and therefore allows the
identification of the states with those of the CFT Hilbert space. The
calculation of this limit can be done in an analytic way by resorting
to the so called {\em kink} DdV equations. Let
$x=\log(r/2)$. The UV limit is then given by
$x\to-\infty$. The shape of the function $-if(\theta)$ in the region
$x\ll 0$, is almost flat
(and its value very close to $\alpha$) inside the region $|\theta|\ll -x$ while
it has a sharp $e^{r\theta}$ behaviour for $\theta\gg -x$ or
$-e^{r\theta}$ for $\theta\ll x$. The hole positions $\theta_h$ are
given by the condition $-if(\theta_h)=2\pi I_h$. For $x\to-\infty$
therefore, $\theta_h\to\mbox{\rm sign}(I_h)\infty$. Let us divide the set of
holes into those with positive rapidity ($I_h>0$) and those with
negative rapidity ($I_h<0$). Call these two sets $\cI_+$ and $\cI_-$
respectively. We assume for simplicity that these two sets contain
half of the holes each, $|\cI_{\pm}|=H/2$. (Remember that if the number
of sites of the lattice is even, the number of holes also must be even).
The most general case will be treated in~\cite{preparation}. 
It is obvious that towards the UV limit, the holes in
$\cI_+$ (resp. $\cI_-$) behave as right (resp. left) kinks. Their
rapidities can be shifted as follows: $\rho_h = \theta_h \pm x$ if
$h\in\cI_{\pm}$. The DdV equation splits in the UV limit into two
right ($+$) and left ($-$) {\em kink} DdV equations, correspondingly to the
general shift $\theta\pm x=\rho$. Dropping terms exponentially small
when $x\to-\infty$ and introducing the two kink pseudoenergies
$f_{\pm}(\rho)\equiv \lim_{x\to-\infty} f(\pm(\rho-x))$, these two 
equations turn out to be scale invariant, i.e. indipendent of $x$
\bea
\mp i f_{\pm}(\rho)&=&e^{\pm\rho} \pm\alpha
+\sfrac{H}{2}\alpha'\mp i\sum_{h\in\cI_{\pm}}\log S(\rho-\rho_h) \nonumber\\
&\pm& 2 \intr
d\rho' G(\rho-\rho') \im \log (1+e^{f_{\pm}(\rho+i0)})
\eea
where we have used the fact that $\log S(\pm\infty)=\pm i\alpha'$ with
$\alpha'=\pi\frac{p-1}{2p}$. Correspondingly, the scaling function
assumes the form $\ct(0)=c_+ + c_-$, with
\eq
c_{\pm}=-\frac{6}{\pi}\left[ \sum_{h\in\cI_{\pm}}e^{\pm\rho_h}
\mp\frac{1}{\pi} \intr d\rho e^{\pm\rho}\im
\log(1+e^{f_{\pm}(\rho+i0)})\right]
\label{uno}
\en 
Introducing $\varphi_{\pm}(\rho)\equiv
e^{\pm\rho}\pm\alpha+\sfrac{H}{2}\alpha' \mp i \sum_{h\in\cI_{\pm}}
\log S(\rho-\rho_h)$ we can write eq.(\ref{uno}) as
\bea
c_{\pm}&=&-\frac{6}{\pi}\sum_{h\in\cI_{\pm}}e^{\pm\rho_h}+\frac{6}{\pi^2}\intr
d\rho\varphi_{\pm}'(\rho)\im\log(1+e^{f_{\pm}(\rho+i0)})\nonumber\\
&\pm&\frac{6}{\pi^2}\sum_{h\in\cI_{\pm}}\intr d\rho i\frac{d}{d\rho}\log
S(\rho-\rho_h) \im\log(1+e^{f_{\pm}(\rho+i0)})
\eea
The second term on the r.h.s. of this equation can be treated by
resorting to the lemma cited in ref.~\cite{DdV2} and observing that 
$f_{\pm}(\mp\infty+i0)=\pm i\alpha\frac{2p}{p+1}$. The first and third
terms can be instead manipulated substituting 
$e^{\pm\rho_h}$ by use of the DdV equation and remebering that 
$\rho_h-\rho_{h'}\to 0$ for
$x\to-\infty$ if both $h,h'$ are in $\cI_+$ or $\cI_-$ and $\log S(0)=0$.
One arrives at the final formula
\eq
\ct(0) = 1 - 12(\Delta_+ + \Delta_-)
\en
with
\eq
\Delta_{\pm}=\frac{(\frac{H}{2}(p+1)\mp \kappa)^2}{4p(p+1)}+K_{\pm}
\virg K_{\pm}=\pm\sum_{h\in\cI_{\pm}}I_h - \frac{H^2}{8}
\en
giving the left and right dimensions of the conformal state
which is the UV limit of the $H$ hole state we are examining. The
(Lorentz) spin of the state is given by
$\Delta_+-\Delta_-=-\frac{H\kappa}{2p}+\sum_{h=1}^H I_h$. The numbers
$K_{\pm}$ are non-negative integers, and account for secondaries. 

The states we are studying are uniquely defined by giving the sequence
of $(I_1,I_2,...,I_H)$. Repetitions of values of $I_h$ must be avoided
in a sequence.
As an example, we discuss here the case
$p=3$. The $\kappa=0$ sector corresponds, at the UV limit, to
conformal states selected by periodic boundary conditions. They are
encoded in the modular invariant partition function of the model. An
inspection of this latter shows that it contains a certain number of
states of spin 0, among which there are the
$\Delta_+=\Delta_-=\frac{1}{3},\frac{4}{3},3$
states. These can be obtained in our framework by exciting 2,4,6 holes
respectively, with the lowest possible values of $|I_h|$
\eq
(-\half,\half) \to \sfrac{1}{3}\virg
(-\sfrac{3}{2},-\half,\half,\sfrac{3}{2}) \to \sfrac{4}{3} \virg
(-\sfrac{5}{2},-\sfrac{3}{2},-\half,\half,\sfrac{3}{2},\sfrac{5}{2})
\to 3
\en
Using larger values of $|I_h|$, one can get secondaries of these
states, even with spin different from zero. Often, there are many
degenerate secondaries with the same values of $\Delta_++\Delta_-$. We
can also obtain the same values of $\Delta_++\Delta_-$ by different
sequences of $I_h$'s. In the example above 2-hole states and 4-hole
states often degenerate in the conformal limit, although their
infrared destiny is very different, going to 2-particle and
4-particle states respectively. To link the description of this
degenerations in terms of hole states with the Virasoro (or extended
conformal algebra) description is a very interesting problem to investigate.

Of course the modular invariant partition function mentioned before,
or its analogs for other values of $p$, contain other states of spin 0
that we cannot reach with our construction as far as we do not deal
with string excitations (see~\cite{preparation}). The hope is, once
both hole and string states are taken into account, to show that the
whole modular invariant partition function can be reconstructed.

If one considers now hole excitations over the other $\alpha$-vacua
with $\kappa\not= 0$, one should expect to reconstruct, in the UV
limit, the partition functions with $\alpha$-twisted boundary
conditions. It is a straightforward exercise of CFT to write down
these partition functions and compare the operator content of these
sectors with our results. The states we find by choosing
2,4,6... holes with the ``minimal'' choice of $I_h$'s have now
fractional spins. It is a matter of fact that such exotic states are
indeed present in the twisted partition functions.

\section{Numerical work}

The results of the previous section have been checked by numerical
integration of the DdV equations. Apart the already mentioned
numerical study of the $\alpha$-vacua, that confirms with high
precision the expected identification with RSG$_p$ excited states on
the diagonal of the Kac-table -- a new result by itself -- we have
investigated numerically the hole excitations with 2,4,6 holes for
values of $p=3,4,5$. The precision, in this investigation has been
taken a bit lower, say 6 or 7 significant digits. It is a simple
choice of speed, no technical problem prevents to get higher precision
data. A detailed summary of these numerical results,
together with their particle interpretation, level crossing phenomena,
etc... will be given
in~\cite{preparation}. Here we only present, as an example of the
method, the energy evolution against $R$ of few of the two and four
holes spin zero excited states over the $\alpha=0$ vacuum in $p=3$. The self
explanatory Fig.1 collects these data. One can clearly see that the
$H$-hole states accumulate to $H$-particle states in the $R\to\infty$
limit, as commented before.

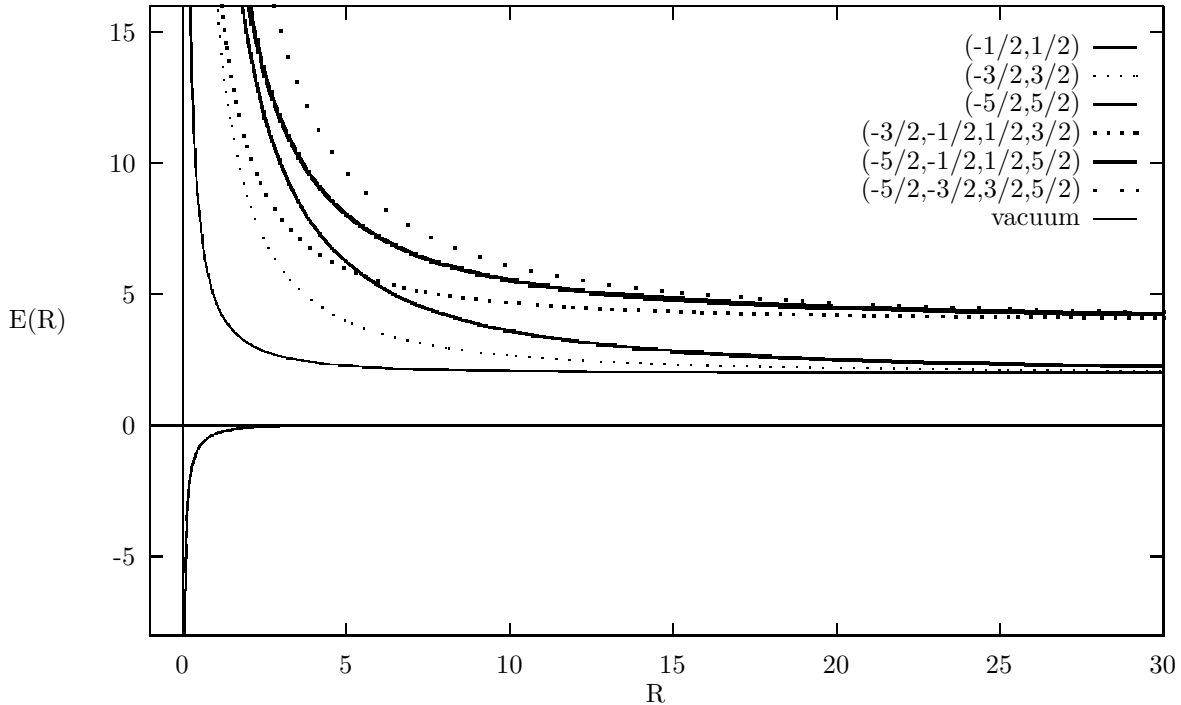
\begin{figure}
\begin{center}
\setlength{\unitlength}{0.240900pt}
\ifx\plotpoint\undefined\newsavebox{\plotpoint}\fi
\sbox{\plotpoint}{\rule[-0.200pt]{0.400pt}{0.400pt}}%
\begin{picture}(1875,1125)(0,0)
\font\gnuplot=cmr10 at 10pt
\gnuplot
\sbox{\plotpoint}{\rule[-0.200pt]{0.400pt}{0.400pt}}%
\put(220.0,443.0){\rule[-0.200pt]{383.272pt}{0.400pt}}
\put(271.0,113.0){\rule[-0.200pt]{0.400pt}{238.250pt}}
\put(220.0,237.0){\rule[-0.200pt]{4.818pt}{0.400pt}}
\put(198,237){\makebox(0,0)[r]{-5}}
\put(1791.0,237.0){\rule[-0.200pt]{4.818pt}{0.400pt}}
\put(220.0,443.0){\rule[-0.200pt]{4.818pt}{0.400pt}}
\put(198,443){\makebox(0,0)[r]{0}}
\put(1791.0,443.0){\rule[-0.200pt]{4.818pt}{0.400pt}}
\put(220.0,649.0){\rule[-0.200pt]{4.818pt}{0.400pt}}
\put(198,649){\makebox(0,0)[r]{5}}
\put(1791.0,649.0){\rule[-0.200pt]{4.818pt}{0.400pt}}
\put(220.0,855.0){\rule[-0.200pt]{4.818pt}{0.400pt}}
\put(198,855){\makebox(0,0)[r]{10}}
\put(1791.0,855.0){\rule[-0.200pt]{4.818pt}{0.400pt}}
\put(220.0,1061.0){\rule[-0.200pt]{4.818pt}{0.400pt}}
\put(198,1061){\makebox(0,0)[r]{15}}
\put(1791.0,1061.0){\rule[-0.200pt]{4.818pt}{0.400pt}}
\put(271.0,113.0){\rule[-0.200pt]{0.400pt}{4.818pt}}
\put(271,68){\makebox(0,0){0}}
\put(271.0,1082.0){\rule[-0.200pt]{0.400pt}{4.818pt}}
\put(528.0,113.0){\rule[-0.200pt]{0.400pt}{4.818pt}}
\put(528,68){\makebox(0,0){5}}
\put(528.0,1082.0){\rule[-0.200pt]{0.400pt}{4.818pt}}
\put(785.0,113.0){\rule[-0.200pt]{0.400pt}{4.818pt}}
\put(785,68){\makebox(0,0){10}}
\put(785.0,1082.0){\rule[-0.200pt]{0.400pt}{4.818pt}}
\put(1041.0,113.0){\rule[-0.200pt]{0.400pt}{4.818pt}}
\put(1041,68){\makebox(0,0){15}}
\put(1041.0,1082.0){\rule[-0.200pt]{0.400pt}{4.818pt}}
\put(1298.0,113.0){\rule[-0.200pt]{0.400pt}{4.818pt}}
\put(1298,68){\makebox(0,0){20}}
\put(1298.0,1082.0){\rule[-0.200pt]{0.400pt}{4.818pt}}
\put(1554.0,113.0){\rule[-0.200pt]{0.400pt}{4.818pt}}
\put(1554,68){\makebox(0,0){25}}
\put(1554.0,1082.0){\rule[-0.200pt]{0.400pt}{4.818pt}}
\put(1811.0,113.0){\rule[-0.200pt]{0.400pt}{4.818pt}}
\put(1811,68){\makebox(0,0){30}}
\put(1811.0,1082.0){\rule[-0.200pt]{0.400pt}{4.818pt}}
\put(220.0,113.0){\rule[-0.200pt]{383.272pt}{0.400pt}}
\put(1811.0,113.0){\rule[-0.200pt]{0.400pt}{238.250pt}}
\put(220.0,1102.0){\rule[-0.200pt]{383.272pt}{0.400pt}}
\put(45,607){\makebox(0,0){E(R)}}
\put(1015,23){\makebox(0,0){R}}
\put(220.0,113.0){\rule[-0.200pt]{0.400pt}{238.250pt}}
\put(1681,1037){\makebox(0,0)[r]{(-1/2,1/2)}}
\put(283.67,1056){\rule{0.400pt}{11.081pt}}
\multiput(283.17,1079.00)(1.000,-23.000){2}{\rule{0.400pt}{5.541pt}}
\put(285.17,982){\rule{0.400pt}{14.900pt}}
\multiput(284.17,1025.07)(2.000,-43.074){2}{\rule{0.400pt}{7.450pt}}
\put(287.17,918){\rule{0.400pt}{12.900pt}}
\multiput(286.17,955.23)(2.000,-37.225){2}{\rule{0.400pt}{6.450pt}}
\multiput(289.61,886.59)(0.447,-12.295){3}{\rule{0.108pt}{7.567pt}}
\multiput(288.17,902.30)(3.000,-40.295){2}{\rule{0.400pt}{3.783pt}}
\multiput(292.61,834.46)(0.447,-10.732){3}{\rule{0.108pt}{6.633pt}}
\multiput(291.17,848.23)(3.000,-35.232){2}{\rule{0.400pt}{3.317pt}}
\multiput(295.60,795.57)(0.468,-5.891){5}{\rule{0.113pt}{4.200pt}}
\multiput(294.17,804.28)(4.000,-32.283){2}{\rule{0.400pt}{2.100pt}}
\multiput(299.60,756.23)(0.468,-5.306){5}{\rule{0.113pt}{3.800pt}}
\multiput(298.17,764.11)(4.000,-29.113){2}{\rule{0.400pt}{1.900pt}}
\multiput(303.60,721.72)(0.468,-4.429){5}{\rule{0.113pt}{3.200pt}}
\multiput(302.17,728.36)(4.000,-24.358){2}{\rule{0.400pt}{1.600pt}}
\multiput(307.59,696.11)(0.482,-2.389){9}{\rule{0.116pt}{1.900pt}}
\multiput(306.17,700.06)(6.000,-23.056){2}{\rule{0.400pt}{0.950pt}}
\multiput(313.59,669.94)(0.482,-2.118){9}{\rule{0.116pt}{1.700pt}}
\multiput(312.17,673.47)(6.000,-20.472){2}{\rule{0.400pt}{0.850pt}}
\multiput(319.59,647.84)(0.485,-1.484){11}{\rule{0.117pt}{1.243pt}}
\multiput(318.17,650.42)(7.000,-17.420){2}{\rule{0.400pt}{0.621pt}}
\multiput(326.59,629.06)(0.488,-1.088){13}{\rule{0.117pt}{0.950pt}}
\multiput(325.17,631.03)(8.000,-15.028){2}{\rule{0.400pt}{0.475pt}}
\multiput(334.58,613.09)(0.491,-0.756){17}{\rule{0.118pt}{0.700pt}}
\multiput(333.17,614.55)(10.000,-13.547){2}{\rule{0.400pt}{0.350pt}}
\multiput(344.58,598.62)(0.492,-0.590){19}{\rule{0.118pt}{0.573pt}}
\multiput(343.17,599.81)(11.000,-11.811){2}{\rule{0.400pt}{0.286pt}}
\multiput(355.00,586.92)(0.543,-0.492){19}{\rule{0.536pt}{0.118pt}}
\multiput(355.00,587.17)(10.887,-11.000){2}{\rule{0.268pt}{0.400pt}}
\multiput(367.00,575.93)(0.844,-0.489){15}{\rule{0.767pt}{0.118pt}}
\multiput(367.00,576.17)(13.409,-9.000){2}{\rule{0.383pt}{0.400pt}}
\multiput(382.00,566.93)(1.022,-0.488){13}{\rule{0.900pt}{0.117pt}}
\multiput(382.00,567.17)(14.132,-8.000){2}{\rule{0.450pt}{0.400pt}}
\multiput(398.00,558.93)(1.666,-0.482){9}{\rule{1.367pt}{0.116pt}}
\multiput(398.00,559.17)(16.163,-6.000){2}{\rule{0.683pt}{0.400pt}}
\multiput(417.00,552.93)(1.937,-0.482){9}{\rule{1.567pt}{0.116pt}}
\multiput(417.00,553.17)(18.748,-6.000){2}{\rule{0.783pt}{0.400pt}}
\multiput(439.00,546.94)(3.698,-0.468){5}{\rule{2.700pt}{0.113pt}}
\multiput(439.00,547.17)(20.396,-4.000){2}{\rule{1.350pt}{0.400pt}}
\multiput(465.00,542.94)(3.990,-0.468){5}{\rule{2.900pt}{0.113pt}}
\multiput(465.00,543.17)(21.981,-4.000){2}{\rule{1.450pt}{0.400pt}}
\multiput(493.00,538.95)(7.383,-0.447){3}{\rule{4.633pt}{0.108pt}}
\multiput(493.00,539.17)(24.383,-3.000){2}{\rule{2.317pt}{0.400pt}}
\multiput(527.00,535.95)(8.276,-0.447){3}{\rule{5.167pt}{0.108pt}}
\multiput(527.00,536.17)(27.276,-3.000){2}{\rule{2.583pt}{0.400pt}}
\put(565,532.17){\rule{8.900pt}{0.400pt}}
\multiput(565.00,533.17)(25.528,-2.000){2}{\rule{4.450pt}{0.400pt}}
\put(609,530.67){\rule{12.286pt}{0.400pt}}
\multiput(609.00,531.17)(25.500,-1.000){2}{\rule{6.143pt}{0.400pt}}
\put(660,529.67){\rule{13.972pt}{0.400pt}}
\multiput(660.00,530.17)(29.000,-1.000){2}{\rule{6.986pt}{0.400pt}}
\put(718,528.67){\rule{16.140pt}{0.400pt}}
\multiput(718.00,529.17)(33.500,-1.000){2}{\rule{8.070pt}{0.400pt}}
\put(785,527.67){\rule{18.549pt}{0.400pt}}
\multiput(785.00,528.17)(38.500,-1.000){2}{\rule{9.275pt}{0.400pt}}
\put(862,526.67){\rule{21.440pt}{0.400pt}}
\multiput(862.00,527.17)(44.500,-1.000){2}{\rule{10.720pt}{0.400pt}}
\put(1703.0,1037.0){\rule[-0.200pt]{15.899pt}{0.400pt}}
\put(1053,525.67){\rule{28.185pt}{0.400pt}}
\multiput(1053.00,526.17)(58.500,-1.000){2}{\rule{14.093pt}{0.400pt}}
\put(951.0,527.0){\rule[-0.200pt]{24.572pt}{0.400pt}}
\put(1170.0,526.0){\rule[-0.200pt]{154.417pt}{0.400pt}}
\put(1681,992){\makebox(0,0)[r]{(-3/2,3/2)}}
\multiput(1703,992)(20.756,0.000){4}{\usebox{\plotpoint}}
\put(1769,992){\usebox{\plotpoint}}
\put(325.00,1102.00){\usebox{\plotpoint}}
\multiput(326,1095)(1.824,-20.675){3}{\usebox{\plotpoint}}
\multiput(332,1027)(2.405,-20.616){3}{\usebox{\plotpoint}}
\multiput(339,967)(3.412,-20.473){3}{\usebox{\plotpoint}}
\multiput(348,913)(3.904,-20.385){2}{\usebox{\plotpoint}}
\multiput(357,866)(4.701,-20.216){2}{\usebox{\plotpoint}}
\multiput(367,823)(5.915,-19.895){2}{\usebox{\plotpoint}}
\multiput(378,786)(7.413,-19.387){2}{\usebox{\plotpoint}}
\put(398.20,737.59){\usebox{\plotpoint}}
\multiput(406,722)(10.581,-17.856){2}{\usebox{\plotpoint}}
\put(430.29,684.41){\usebox{\plotpoint}}
\multiput(440,672)(14.314,-15.030){2}{\usebox{\plotpoint}}
\put(473.79,640.21){\usebox{\plotpoint}}
\put(490.63,628.12){\usebox{\plotpoint}}
\multiput(508,617)(18.691,-9.023){2}{\usebox{\plotpoint}}
\multiput(537,603)(19.356,-7.493){2}{\usebox{\plotpoint}}
\put(585.07,586.26){\usebox{\plotpoint}}
\multiput(604,581)(20.249,-4.556){2}{\usebox{\plotpoint}}
\multiput(644,572)(20.435,-3.633){3}{\usebox{\plotpoint}}
\multiput(689,564)(20.555,-2.878){2}{\usebox{\plotpoint}}
\multiput(739,557)(20.673,-1.846){3}{\usebox{\plotpoint}}
\multiput(795,552)(20.690,-1.642){3}{\usebox{\plotpoint}}
\multiput(858,547)(20.722,-1.184){3}{\usebox{\plotpoint}}
\multiput(928,543)(20.741,-0.788){4}{\usebox{\plotpoint}}
\multiput(1007,540)(20.743,-0.707){4}{\usebox{\plotpoint}}
\multiput(1095,537)(20.751,-0.419){5}{\usebox{\plotpoint}}
\multiput(1194,535)(20.752,-0.374){5}{\usebox{\plotpoint}}
\multiput(1305,533)(20.755,-0.167){6}{\usebox{\plotpoint}}
\multiput(1429,532)(20.753,-0.301){7}{\usebox{\plotpoint}}
\multiput(1567,530)(20.755,-0.133){7}{\usebox{\plotpoint}}
\multiput(1723,529)(20.756,0.000){5}{\usebox{\plotpoint}}
\put(1811,529){\usebox{\plotpoint}}
\sbox{\plotpoint}{\rule[-0.400pt]{0.800pt}{0.800pt}}%
\put(1681,947){\makebox(0,0)[r]{(-5/2,5/2)}}
\multiput(367.38,1079.25)(0.560,-5.126){3}{\rule{0.135pt}{5.480pt}}
\multiput(364.34,1090.63)(5.000,-21.626){2}{\rule{0.800pt}{2.740pt}}
\multiput(372.40,1049.72)(0.516,-3.072){11}{\rule{0.124pt}{4.644pt}}
\multiput(369.34,1059.36)(9.000,-40.360){2}{\rule{0.800pt}{2.322pt}}
\multiput(381.40,1002.89)(0.514,-2.491){13}{\rule{0.124pt}{3.880pt}}
\multiput(378.34,1010.95)(10.000,-37.947){2}{\rule{0.800pt}{1.940pt}}
\multiput(391.40,957.89)(0.514,-2.324){13}{\rule{0.124pt}{3.640pt}}
\multiput(388.34,965.44)(10.000,-35.445){2}{\rule{0.800pt}{1.820pt}}
\multiput(401.41,918.65)(0.511,-1.666){17}{\rule{0.123pt}{2.733pt}}
\multiput(398.34,924.33)(12.000,-32.327){2}{\rule{0.800pt}{1.367pt}}
\multiput(413.41,882.23)(0.509,-1.402){19}{\rule{0.123pt}{2.354pt}}
\multiput(410.34,887.11)(13.000,-30.114){2}{\rule{0.800pt}{1.177pt}}
\multiput(426.41,848.34)(0.509,-1.217){21}{\rule{0.123pt}{2.086pt}}
\multiput(423.34,852.67)(14.000,-28.671){2}{\rule{0.800pt}{1.043pt}}
\multiput(440.41,816.53)(0.508,-1.024){23}{\rule{0.122pt}{1.800pt}}
\multiput(437.34,820.26)(15.000,-26.264){2}{\rule{0.800pt}{0.900pt}}
\multiput(455.41,787.57)(0.507,-0.857){25}{\rule{0.122pt}{1.550pt}}
\multiput(452.34,790.78)(16.000,-23.783){2}{\rule{0.800pt}{0.775pt}}
\multiput(471.41,761.74)(0.506,-0.669){29}{\rule{0.122pt}{1.267pt}}
\multiput(468.34,764.37)(18.000,-21.371){2}{\rule{0.800pt}{0.633pt}}
\multiput(489.41,738.15)(0.506,-0.605){31}{\rule{0.122pt}{1.168pt}}
\multiput(486.34,740.57)(19.000,-20.575){2}{\rule{0.800pt}{0.584pt}}
\multiput(507.00,718.09)(0.547,-0.505){33}{\rule{1.080pt}{0.122pt}}
\multiput(507.00,718.34)(19.758,-20.000){2}{\rule{0.540pt}{0.800pt}}
\multiput(529.00,698.09)(0.605,-0.506){31}{\rule{1.168pt}{0.122pt}}
\multiput(529.00,698.34)(20.575,-19.000){2}{\rule{0.584pt}{0.800pt}}
\multiput(552.00,679.09)(0.742,-0.507){27}{\rule{1.376pt}{0.122pt}}
\multiput(552.00,679.34)(22.143,-17.000){2}{\rule{0.688pt}{0.800pt}}
\multiput(577.00,662.09)(0.857,-0.507){25}{\rule{1.550pt}{0.122pt}}
\multiput(577.00,662.34)(23.783,-16.000){2}{\rule{0.775pt}{0.800pt}}
\multiput(604.00,646.09)(1.103,-0.509){21}{\rule{1.914pt}{0.123pt}}
\multiput(604.00,646.34)(26.027,-14.000){2}{\rule{0.957pt}{0.800pt}}
\multiput(634.00,632.08)(1.440,-0.511){17}{\rule{2.400pt}{0.123pt}}
\multiput(634.00,632.34)(28.019,-12.000){2}{\rule{1.200pt}{0.800pt}}
\multiput(667.00,620.08)(1.736,-0.512){15}{\rule{2.818pt}{0.123pt}}
\multiput(667.00,620.34)(30.151,-11.000){2}{\rule{1.409pt}{0.800pt}}
\multiput(703.00,609.08)(1.885,-0.512){15}{\rule{3.036pt}{0.123pt}}
\multiput(703.00,609.34)(32.698,-11.000){2}{\rule{1.518pt}{0.800pt}}
\multiput(742.00,598.08)(2.567,-0.516){11}{\rule{3.933pt}{0.124pt}}
\multiput(742.00,598.34)(33.836,-9.000){2}{\rule{1.967pt}{0.800pt}}
\multiput(784.00,589.08)(3.259,-0.520){9}{\rule{4.800pt}{0.125pt}}
\multiput(784.00,589.34)(36.037,-8.000){2}{\rule{2.400pt}{0.800pt}}
\multiput(830.00,581.08)(4.242,-0.526){7}{\rule{5.914pt}{0.127pt}}
\multiput(830.00,581.34)(37.725,-7.000){2}{\rule{2.957pt}{0.800pt}}
\multiput(880.00,574.08)(4.680,-0.526){7}{\rule{6.486pt}{0.127pt}}
\multiput(880.00,574.34)(41.539,-7.000){2}{\rule{3.243pt}{0.800pt}}
\multiput(935.00,567.07)(6.490,-0.536){5}{\rule{8.200pt}{0.129pt}}
\multiput(935.00,567.34)(42.980,-6.000){2}{\rule{4.100pt}{0.800pt}}
\multiput(995.00,561.06)(10.499,-0.560){3}{\rule{10.600pt}{0.135pt}}
\multiput(995.00,561.34)(42.999,-5.000){2}{\rule{5.300pt}{0.800pt}}
\put(1060,554.34){\rule{14.400pt}{0.800pt}}
\multiput(1060.00,556.34)(41.112,-4.000){2}{\rule{7.200pt}{0.800pt}}
\put(1131,550.34){\rule{15.600pt}{0.800pt}}
\multiput(1131.00,552.34)(44.621,-4.000){2}{\rule{7.800pt}{0.800pt}}
\put(1208,546.34){\rule{17.200pt}{0.800pt}}
\multiput(1208.00,548.34)(49.301,-4.000){2}{\rule{8.600pt}{0.800pt}}
\put(1293,542.84){\rule{22.163pt}{0.800pt}}
\multiput(1293.00,544.34)(46.000,-3.000){2}{\rule{11.081pt}{0.800pt}}
\put(1385,540.34){\rule{24.090pt}{0.800pt}}
\multiput(1385.00,541.34)(50.000,-2.000){2}{\rule{12.045pt}{0.800pt}}
\put(1485,537.84){\rule{26.258pt}{0.800pt}}
\multiput(1485.00,539.34)(54.500,-3.000){2}{\rule{13.129pt}{0.800pt}}
\put(1594,535.84){\rule{28.667pt}{0.800pt}}
\multiput(1594.00,536.34)(59.500,-1.000){2}{\rule{14.334pt}{0.800pt}}
\put(1713,534.34){\rule{23.608pt}{0.800pt}}
\multiput(1713.00,535.34)(49.000,-2.000){2}{\rule{11.804pt}{0.800pt}}
\put(1703.0,947.0){\rule[-0.400pt]{15.899pt}{0.800pt}}
\sbox{\plotpoint}{\rule[-0.500pt]{1.000pt}{1.000pt}}%
\put(1681,902){\makebox(0,0)[r]{(-3/2,-1/2,1/2,3/2)}}
\multiput(1703,902)(20.756,0.000){4}{\usebox{\plotpoint}}
\put(1769,902){\usebox{\plotpoint}}
\put(332.00,1102.00){\usebox{\plotpoint}}
\multiput(333,1085)(2.935,-20.547){3}{\usebox{\plotpoint}}
\multiput(340,1036)(2.743,-20.573){2}{\usebox{\plotpoint}}
\multiput(346,991)(4.070,-20.352){2}{\usebox{\plotpoint}}
\multiput(354,951)(4.386,-20.287){2}{\usebox{\plotpoint}}
\put(366.47,897.62){\usebox{\plotpoint}}
\multiput(371,881)(6.563,-19.690){2}{\usebox{\plotpoint}}
\put(386.06,838.57){\usebox{\plotpoint}}
\multiput(392,824)(9.282,-18.564){2}{\usebox{\plotpoint}}
\put(414.58,783.38){\usebox{\plotpoint}}
\put(426.25,766.22){\usebox{\plotpoint}}
\put(439.58,750.42){\usebox{\plotpoint}}
\put(454.41,735.90){\usebox{\plotpoint}}
\put(469.88,722.09){\usebox{\plotpoint}}
\multiput(486,710)(18.021,-10.298){2}{\usebox{\plotpoint}}
\put(523.30,690.53){\usebox{\plotpoint}}
\put(542.31,682.22){\usebox{\plotpoint}}
\multiput(557,676)(19.760,-6.351){2}{\usebox{\plotpoint}}
\put(601.40,662.90){\usebox{\plotpoint}}
\multiput(617,659)(20.329,-4.185){2}{\usebox{\plotpoint}}
\multiput(651,652)(20.502,-3.237){2}{\usebox{\plotpoint}}
\multiput(689,646)(20.610,-2.454){2}{\usebox{\plotpoint}}
\multiput(731,641)(20.634,-2.243){2}{\usebox{\plotpoint}}
\multiput(777,636)(20.689,-1.655){2}{\usebox{\plotpoint}}
\multiput(827,632)(20.703,-1.479){3}{\usebox{\plotpoint}}
\multiput(883,628)(20.730,-1.020){3}{\usebox{\plotpoint}}
\multiput(944,625)(20.746,-0.619){3}{\usebox{\plotpoint}}
\multiput(1011,623)(20.738,-0.841){4}{\usebox{\plotpoint}}
\multiput(1085,620)(20.749,-0.506){4}{\usebox{\plotpoint}}
\multiput(1167,618)(20.754,-0.233){4}{\usebox{\plotpoint}}
\multiput(1256,617)(20.751,-0.419){5}{\usebox{\plotpoint}}
\multiput(1355,615)(20.755,-0.192){5}{\usebox{\plotpoint}}
\multiput(1463,614)(20.755,-0.174){6}{\usebox{\plotpoint}}
\multiput(1582,613)(20.755,-0.157){6}{\usebox{\plotpoint}}
\multiput(1714,612)(20.756,0.000){5}{\usebox{\plotpoint}}
\put(1811,612){\usebox{\plotpoint}}
\sbox{\plotpoint}{\rule[-0.600pt]{1.200pt}{1.200pt}}%
\put(1681,857){\makebox(0,0)[r]{(-5/2,-1/2,1/2,5/2)}}
\put(376.01,1084){\rule{1.200pt}{4.336pt}}
\multiput(374.51,1093.00)(3.000,-9.000){2}{\rule{1.200pt}{2.168pt}}
\multiput(382.24,1060.34)(0.502,-2.371){10}{\rule{0.121pt}{5.700pt}}
\multiput(377.51,1072.17)(10.000,-33.169){2}{\rule{1.200pt}{2.850pt}}
\multiput(392.24,1017.33)(0.502,-2.148){10}{\rule{0.121pt}{5.220pt}}
\multiput(387.51,1028.17)(10.000,-30.166){2}{\rule{1.200pt}{2.610pt}}
\multiput(402.24,980.98)(0.501,-1.614){14}{\rule{0.121pt}{4.100pt}}
\multiput(397.51,989.49)(12.000,-29.490){2}{\rule{1.200pt}{2.050pt}}
\multiput(414.24,945.73)(0.501,-1.315){16}{\rule{0.121pt}{3.438pt}}
\multiput(409.51,952.86)(13.000,-26.863){2}{\rule{1.200pt}{1.719pt}}
\multiput(427.24,913.37)(0.501,-1.139){18}{\rule{0.121pt}{3.043pt}}
\multiput(422.51,919.68)(14.000,-25.684){2}{\rule{1.200pt}{1.521pt}}
\multiput(441.24,883.46)(0.501,-0.919){20}{\rule{0.121pt}{2.540pt}}
\multiput(436.51,888.73)(15.000,-22.728){2}{\rule{1.200pt}{1.270pt}}
\multiput(456.24,856.35)(0.501,-0.826){22}{\rule{0.121pt}{2.325pt}}
\multiput(451.51,861.17)(16.000,-22.174){2}{\rule{1.200pt}{1.163pt}}
\multiput(472.24,831.39)(0.501,-0.616){26}{\rule{0.121pt}{1.833pt}}
\multiput(467.51,835.19)(18.000,-19.195){2}{\rule{1.200pt}{0.917pt}}
\multiput(490.24,808.99)(0.501,-0.555){28}{\rule{0.121pt}{1.689pt}}
\multiput(485.51,812.49)(19.000,-18.493){2}{\rule{1.200pt}{0.845pt}}
\multiput(507.00,791.26)(0.527,-0.501){30}{\rule{1.620pt}{0.121pt}}
\multiput(507.00,791.51)(18.638,-20.000){2}{\rule{0.810pt}{1.200pt}}
\multiput(529.00,771.26)(0.616,-0.501){26}{\rule{1.833pt}{0.121pt}}
\multiput(529.00,771.51)(19.195,-18.000){2}{\rule{0.917pt}{1.200pt}}
\multiput(552.00,753.26)(0.761,-0.501){22}{\rule{2.175pt}{0.121pt}}
\multiput(552.00,753.51)(20.486,-16.000){2}{\rule{1.088pt}{1.200pt}}
\multiput(577.00,737.26)(0.950,-0.501){18}{\rule{2.614pt}{0.121pt}}
\multiput(577.00,737.51)(21.574,-14.000){2}{\rule{1.307pt}{1.200pt}}
\multiput(604.00,723.26)(1.064,-0.501){18}{\rule{2.871pt}{0.121pt}}
\multiput(604.00,723.51)(24.040,-14.000){2}{\rule{1.436pt}{1.200pt}}
\multiput(634.00,709.26)(1.389,-0.501){14}{\rule{3.600pt}{0.121pt}}
\multiput(634.00,709.51)(25.528,-12.000){2}{\rule{1.800pt}{1.200pt}}
\multiput(667.00,697.26)(1.869,-0.502){10}{\rule{4.620pt}{0.121pt}}
\multiput(667.00,697.51)(26.411,-10.000){2}{\rule{2.310pt}{1.200pt}}
\multiput(703.00,687.26)(2.036,-0.502){10}{\rule{4.980pt}{0.121pt}}
\multiput(703.00,687.51)(28.664,-10.000){2}{\rule{2.490pt}{1.200pt}}
\multiput(742.00,677.26)(2.501,-0.502){8}{\rule{5.900pt}{0.121pt}}
\multiput(742.00,677.51)(29.754,-9.000){2}{\rule{2.950pt}{1.200pt}}
\multiput(784.00,668.26)(3.222,-0.503){6}{\rule{7.200pt}{0.121pt}}
\multiput(784.00,668.51)(31.056,-8.000){2}{\rule{3.600pt}{1.200pt}}
\multiput(830.00,660.26)(4.391,-0.505){4}{\rule{8.871pt}{0.122pt}}
\multiput(830.00,660.51)(31.587,-7.000){2}{\rule{4.436pt}{1.200pt}}
\multiput(880.00,653.25)(8.093,-0.509){2}{\rule{11.300pt}{0.123pt}}
\multiput(880.00,653.51)(31.546,-6.000){2}{\rule{5.650pt}{1.200pt}}
\put(935,645.01){\rule{14.454pt}{1.200pt}}
\multiput(935.00,647.51)(30.000,-5.000){2}{\rule{7.227pt}{1.200pt}}
\put(995,640.01){\rule{15.658pt}{1.200pt}}
\multiput(995.00,642.51)(32.500,-5.000){2}{\rule{7.829pt}{1.200pt}}
\put(1060,635.51){\rule{17.104pt}{1.200pt}}
\multiput(1060.00,637.51)(35.500,-4.000){2}{\rule{8.552pt}{1.200pt}}
\put(1131,631.51){\rule{18.549pt}{1.200pt}}
\multiput(1131.00,633.51)(38.500,-4.000){2}{\rule{9.275pt}{1.200pt}}
\put(1208,627.51){\rule{20.476pt}{1.200pt}}
\multiput(1208.00,629.51)(42.500,-4.000){2}{\rule{10.238pt}{1.200pt}}
\put(1293,624.51){\rule{22.163pt}{1.200pt}}
\multiput(1293.00,625.51)(46.000,-2.000){2}{\rule{11.081pt}{1.200pt}}
\put(1385,622.01){\rule{24.090pt}{1.200pt}}
\multiput(1385.00,623.51)(50.000,-3.000){2}{\rule{12.045pt}{1.200pt}}
\put(1485,619.51){\rule{26.258pt}{1.200pt}}
\multiput(1485.00,620.51)(54.500,-2.000){2}{\rule{13.129pt}{1.200pt}}
\put(1594,617.51){\rule{28.667pt}{1.200pt}}
\multiput(1594.00,618.51)(59.500,-2.000){2}{\rule{14.334pt}{1.200pt}}
\put(1713,616.01){\rule{23.608pt}{1.200pt}}
\multiput(1713.00,616.51)(49.000,-1.000){2}{\rule{11.804pt}{1.200pt}}
\put(1703.0,857.0){\rule[-0.600pt]{15.899pt}{1.200pt}}
\sbox{\plotpoint}{\rule[-0.500pt]{1.000pt}{1.000pt}}%
\put(1681,812){\makebox(0,0)[r]{(-5/2,-3/2,3/2,5/2)}}
\multiput(1703,812)(41.511,0.000){2}{\usebox{\plotpoint}}
\put(1769,812){\usebox{\plotpoint}}
\put(413.00,1102.00){\usebox{\plotpoint}}
\put(424.12,1062.01){\usebox{\plotpoint}}
\put(435.81,1022.18){\usebox{\plotpoint}}
\put(449.50,983.01){\usebox{\plotpoint}}
\put(464.77,944.41){\usebox{\plotpoint}}
\put(483.29,907.32){\usebox{\plotpoint}}
\multiput(484,906)(21.162,-35.712){0}{\usebox{\plotpoint}}
\put(504.85,871.86){\usebox{\plotpoint}}
\put(529.68,838.65){\usebox{\plotpoint}}
\multiput(536,831)(28.628,-30.060){0}{\usebox{\plotpoint}}
\put(557.92,808.26){\usebox{\plotpoint}}
\put(589.43,781.27){\usebox{\plotpoint}}
\put(623.37,757.42){\usebox{\plotpoint}}
\multiput(624,757)(36.287,-20.160){0}{\usebox{\plotpoint}}
\put(659.95,737.85){\usebox{\plotpoint}}
\put(698.12,721.60){\usebox{\plotpoint}}
\put(737.04,707.17){\usebox{\plotpoint}}
\put(776.82,695.34){\usebox{\plotpoint}}
\multiput(778,695)(40.394,-9.567){0}{\usebox{\plotpoint}}
\put(817.20,685.74){\usebox{\plotpoint}}
\multiput(857,677)(40.995,-6.522){2}{\usebox{\plotpoint}}
\put(939.81,664.22){\usebox{\plotpoint}}
\put(981.00,659.12){\usebox{\plotpoint}}
\put(1022.25,654.46){\usebox{\plotpoint}}
\multiput(1054,651)(41.358,-3.565){2}{\usebox{\plotpoint}}
\put(1146.31,643.86){\usebox{\plotpoint}}
\multiput(1176,642)(41.437,-2.474){2}{\usebox{\plotpoint}}
\multiput(1243,638)(41.449,-2.271){2}{\usebox{\plotpoint}}
\put(1353.55,632.57){\usebox{\plotpoint}}
\multiput(1395,631)(41.485,-1.482){3}{\usebox{\plotpoint}}
\multiput(1479,628)(41.501,-0.922){2}{\usebox{\plotpoint}}
\multiput(1569,626)(41.492,-1.270){2}{\usebox{\plotpoint}}
\multiput(1667,623)(41.503,-0.798){3}{\usebox{\plotpoint}}
\put(1810.00,621.00){\usebox{\plotpoint}}
\put(1811,621){\usebox{\plotpoint}}
\sbox{\plotpoint}{\rule[-0.200pt]{0.400pt}{0.400pt}}%
\put(1681,767){\makebox(0,0)[r]{vacuum}}
\put(1703.0,767.0){\rule[-0.200pt]{15.899pt}{0.400pt}}
\put(1811,443){\usebox{\plotpoint}}
\put(502,441.67){\rule{6.263pt}{0.400pt}}
\multiput(515.00,442.17)(-13.000,-1.000){2}{\rule{3.132pt}{0.400pt}}
\put(528.0,443.0){\rule[-0.200pt]{309.075pt}{0.400pt}}
\put(408,440.67){\rule{3.614pt}{0.400pt}}
\multiput(415.50,441.17)(-7.500,-1.000){2}{\rule{1.807pt}{0.400pt}}
\put(423.0,442.0){\rule[-0.200pt]{19.031pt}{0.400pt}}
\put(382,439.67){\rule{2.891pt}{0.400pt}}
\multiput(388.00,440.17)(-6.000,-1.000){2}{\rule{1.445pt}{0.400pt}}
\put(371,438.67){\rule{2.650pt}{0.400pt}}
\multiput(376.50,439.17)(-5.500,-1.000){2}{\rule{1.325pt}{0.400pt}}
\put(361,437.67){\rule{2.409pt}{0.400pt}}
\multiput(366.00,438.17)(-5.000,-1.000){2}{\rule{1.204pt}{0.400pt}}
\put(352,436.67){\rule{2.168pt}{0.400pt}}
\multiput(356.50,437.17)(-4.500,-1.000){2}{\rule{1.084pt}{0.400pt}}
\put(344,435.17){\rule{1.700pt}{0.400pt}}
\multiput(348.47,436.17)(-4.472,-2.000){2}{\rule{0.850pt}{0.400pt}}
\put(337,433.67){\rule{1.686pt}{0.400pt}}
\multiput(340.50,434.17)(-3.500,-1.000){2}{\rule{0.843pt}{0.400pt}}
\put(330,432.17){\rule{1.500pt}{0.400pt}}
\multiput(333.89,433.17)(-3.887,-2.000){2}{\rule{0.750pt}{0.400pt}}
\put(324,430.17){\rule{1.300pt}{0.400pt}}
\multiput(327.30,431.17)(-3.302,-2.000){2}{\rule{0.650pt}{0.400pt}}
\multiput(320.82,428.95)(-0.909,-0.447){3}{\rule{0.767pt}{0.108pt}}
\multiput(322.41,429.17)(-3.409,-3.000){2}{\rule{0.383pt}{0.400pt}}
\put(314,425.17){\rule{1.100pt}{0.400pt}}
\multiput(316.72,426.17)(-2.717,-2.000){2}{\rule{0.550pt}{0.400pt}}
\multiput(311.37,423.95)(-0.685,-0.447){3}{\rule{0.633pt}{0.108pt}}
\multiput(312.69,424.17)(-2.685,-3.000){2}{\rule{0.317pt}{0.400pt}}
\multiput(307.92,420.94)(-0.481,-0.468){5}{\rule{0.500pt}{0.113pt}}
\multiput(308.96,421.17)(-2.962,-4.000){2}{\rule{0.250pt}{0.400pt}}
\multiput(303.92,416.95)(-0.462,-0.447){3}{\rule{0.500pt}{0.108pt}}
\multiput(304.96,417.17)(-1.962,-3.000){2}{\rule{0.250pt}{0.400pt}}
\multiput(300.92,413.94)(-0.481,-0.468){5}{\rule{0.500pt}{0.113pt}}
\multiput(301.96,414.17)(-2.962,-4.000){2}{\rule{0.250pt}{0.400pt}}
\put(297.17,406){\rule{0.400pt}{1.100pt}}
\multiput(298.17,408.72)(-2.000,-2.717){2}{\rule{0.400pt}{0.550pt}}
\multiput(295.95,402.82)(-0.447,-0.909){3}{\rule{0.108pt}{0.767pt}}
\multiput(296.17,404.41)(-3.000,-3.409){2}{\rule{0.400pt}{0.383pt}}
\put(292.17,396){\rule{0.400pt}{1.100pt}}
\multiput(293.17,398.72)(-2.000,-2.717){2}{\rule{0.400pt}{0.550pt}}
\put(290.17,389){\rule{0.400pt}{1.500pt}}
\multiput(291.17,392.89)(-2.000,-3.887){2}{\rule{0.400pt}{0.750pt}}
\put(288.17,383){\rule{0.400pt}{1.300pt}}
\multiput(289.17,386.30)(-2.000,-3.302){2}{\rule{0.400pt}{0.650pt}}
\put(286.17,375){\rule{0.400pt}{1.700pt}}
\multiput(287.17,379.47)(-2.000,-4.472){2}{\rule{0.400pt}{0.850pt}}
\put(284.67,367){\rule{0.400pt}{1.927pt}}
\multiput(285.17,371.00)(-1.000,-4.000){2}{\rule{0.400pt}{0.964pt}}
\put(283.17,357){\rule{0.400pt}{2.100pt}}
\multiput(284.17,362.64)(-2.000,-5.641){2}{\rule{0.400pt}{1.050pt}}
\put(281.67,347){\rule{0.400pt}{2.409pt}}
\multiput(282.17,352.00)(-1.000,-5.000){2}{\rule{0.400pt}{1.204pt}}
\put(280.67,336){\rule{0.400pt}{2.650pt}}
\multiput(281.17,341.50)(-1.000,-5.500){2}{\rule{0.400pt}{1.325pt}}
\put(279.67,323){\rule{0.400pt}{3.132pt}}
\multiput(280.17,329.50)(-1.000,-6.500){2}{\rule{0.400pt}{1.566pt}}
\put(278.67,309){\rule{0.400pt}{3.373pt}}
\multiput(279.17,316.00)(-1.000,-7.000){2}{\rule{0.400pt}{1.686pt}}
\put(277.67,293){\rule{0.400pt}{3.854pt}}
\multiput(278.17,301.00)(-1.000,-8.000){2}{\rule{0.400pt}{1.927pt}}
\put(394.0,441.0){\rule[-0.200pt]{3.373pt}{0.400pt}}
\put(276.67,257){\rule{0.400pt}{4.577pt}}
\multiput(277.17,266.50)(-1.000,-9.500){2}{\rule{0.400pt}{2.289pt}}
\put(278.0,276.0){\rule[-0.200pt]{0.400pt}{4.095pt}}
\put(275.67,212){\rule{0.400pt}{5.541pt}}
\multiput(276.17,223.50)(-1.000,-11.500){2}{\rule{0.400pt}{2.770pt}}
\put(277.0,235.0){\rule[-0.200pt]{0.400pt}{5.300pt}}
\put(274.67,156){\rule{0.400pt}{6.986pt}}
\multiput(275.17,170.50)(-1.000,-14.500){2}{\rule{0.400pt}{3.493pt}}
\put(276.0,185.0){\rule[-0.200pt]{0.400pt}{6.504pt}}
\put(275.0,113.0){\rule[-0.200pt]{0.400pt}{10.359pt}}
\end{picture}
\caption{Energy versus $R$ of spin 0 pure hole excitations in the $\alpha=0$
sector of SG$_{p=3}$ model.}
\end{center}
\end{figure}

It is interesting to note the undoutable existence, in this graph, of
a level crossing between the $(-\sfrac{5}{2},\sfrac{5}{2})$ and the
$(-\sfrac{3}{2},-\half,\half,\sfrac{3}{2})$ states at $R=5.613...$. Such
level crossings must be present in an integrable model. One of the
advantages of knowing scaling functions exactly, compared to other
approximate methods of investigation like e.g. TCSA, is that the
existence of a level crossing can be established with no doubt. There
is no danger here to confuse a level crossing with two lines that
approach very close and then repulse each other without crossing:
each line, in the DdV approach, is computed indipendently from the others.

Of course there are many states missing in Fig.1. We have not
investigated all the states that include strings in their Bethe root
configurations. These should complete the picture and most probably
provide for more level crossings. We have also done some investigation
in the other $\alpha$-sectors, that we do not reproduce here for pure
reasons of space. In these sectors the number of spin 0 states is much
lower. It is interesting to notice the existence of fractional spin
states. The $\alpha=\pi$ sector is particularly interesting, because
it realizes the antiperiodic boundary conditions where it should be
possible to see the spin $\half$ states corresponding to the Thirring
fermions.

\section{Conclusions}

We have considered here hole excitations of the SG model over the
$\alpha$-vacua, finding a DdV equation for them and calculating their
exact scaling functions, numerically for generic $R$ and analitically
for $R\to 0$. In this latter limit, we have found agreement with the
operator content dictated by the CFT partition functions in the
variuos twisted sectors. Of course, we have reproduced only a part of
the spectrum, because we have not considered more complicated states
including strings. We intend to deal with the general problem in a forthcoming
publication~\cite{preparation}. In this
short letter we made the choice to deal with this technically simpler
problem to probe the method and show its
efficiency. The only states we have provided for
RSG$_p$ models through q-reduction are for the moment those
corresponding to the alpha vacua, i.e. the off-critical deformations
of the {\em spin} 
primary states which lie on the diagonal of the Kac-table. It
is of course obvious that once the full set of SG states will be
treated, access to the scaling functions for all the RSG$_p$ states
will also be available. The extension of the method to non-integer
rational $p$ could then allow the treatment of non-unitary minimal
models perturbed by $\phi_{13}$ too. Among these, the Lee-Yang model
plays the particular role of being the simplest model, on which both
TCSA data and recent TBA exploration of few excited states are
available~\cite{BLZ,Dorey-Tateo}. 
It would be extremely interesting to make a comparison
between these results and our approach. In particular the method of
ref.~\cite{Dorey-Tateo} suggests to explore monodromy issues of the
scaling functions also in our approach. The link between monodromies and Bethe
ansatz interpretation of the results would certainly provide more
insight in the problem of excited states in general.
\vskip 0.3cm

{\bf Acknowledgements} - We are endebted to F.Capocasa, C.Destri and P.E.Dorey
for useful discussions. G.Vacca is also thanked for interest in this
work. E.Q. thanks the warm hospitality at
Physikalisches Institut der Universit\"at Bonn during the early stages
of this project.
This work was supported in part by NATO Grant CRG 950751.

\end{document}